\documentclass[12pt,english,floatfix,showkeys,superscriptaddress,aps,prd,preprint]{revtex4}
\usepackage[latin1]{inputenc}
\usepackage[T1]{fontenc}
\usepackage{lmodern}
\usepackage{amsmath}
\usepackage{amssymb}
\usepackage{graphicx}
\usepackage{caption}
\usepackage{subcaption}
\usepackage{float}
\usepackage{esint}
\usepackage{dcolumn}
\usepackage{babel}
\usepackage{csquotes}
\usepackage{color}
\usepackage{slashed}
\usepackage{simplewick}
\usepackage{hyperref}
\hypersetup{
    colorlinks,
    citecolor=blue,
    filecolor=green,
    linkcolor=purple,
    urlcolor=red,
}

\usepackage{slashed}

\usepackage{hyperref}
\hypersetup{colorlinks,breaklinks,
			citecolor=[rgb]{0,0.0,1.0},
            urlcolor=[rgb]{0.0,0.0,1.0},
            linkcolor=[rgb]{0,0.5,0.9}}

\begin{document}

\title{Field Sources for Wormholes With Multiple Throats/Anti-throats}
\author{T. M. Crispim}
\email{tiago.crispim@fisica.ufc.br}
\affiliation{Departamento de F\'isica, Universidade Federal do Cear\'a, Caixa Postal 6030,\\ Campus do Pici, 60455-760 Fortaleza, Cear\'a, Brazil.}
\author{Marcos V. de S. Silva\footnote{Author to whom any correspondence should be addressed.}}
\email{marcosvinicius@fisica.ufc.br}
\affiliation{Departamento de F\'isica, Universidade Federal do Cear\'a, Caixa Postal 6030,\\ Campus do Pici, 60455-760 Fortaleza, Cear\'a, Brazil.}
\author{G. Alencar}
\email{geova@fisica.ufc.br}
\affiliation{Departamento de F\'isica, Universidade Federal do Cear\'a, Caixa Postal 6030,\\ Campus do Pici, 60455-760 Fortaleza, Cear\'a, Brazil.}
\author{Celio R. Muniz}
\email{celio.muniz@uece.br}
\affiliation{Universidade Estadual do Cear\'a (UECE), Faculdade de Educa\c{c}\~ao, Ci\^encias e Letras de Iguatu, Av. D\'ario Rabelo s/n, Iguatu - CE, 63.500-00 - Brazil.}
\author{Diego S\'aez-Chill\'on G\'omez}
\email{diego.saez@uva.es} 
\affiliation{Department of Theoretical Physics, Atomic and Optics, and Laboratory for Disruptive Interdisciplinary Science (LaDIS), Campus Miguel Delibes, \\ University of Valladolid UVA, Paseo Bel\'en, 7,
47011 - Valladolid, Spain}
\affiliation{Departamento de F\'isica, Universidade Federal do Cear\'a, Caixa Postal 6030,\\ Campus do Pici, 60455-760 Fortaleza, Cear\'a, Brazil.}


\date{\today}

\begin{abstract}
In this work, we investigate wormhole geometries with multiple throats and anti-throats in general relativity. The existence of these structures is identified through the analysis of minima and maxima in the area of the solution. Using embedding diagrams, we visualize the geometry and demonstrate that these objects exhibit a complex structure, distinct from standard single-throat wormholes. We further analyze the geodesic motion in such spacetimes. The solutions are derived from Einstein's equations by coupling a phantom scalar field to nonlinear electrodynamics, and we show that distinct scalar field profiles can generate the same spacetime geometry. Additionally, we examine the energy conditions and demonstrate that, for specific parameter choices, all energy conditions can be partially satisfied in certain regions of spacetime.
\end{abstract}


\maketitle

\section{Introduction}

Recent observations of black hole shadows by the Event Horizon Telescope (EHT) \cite{EventHorizonTelescope:2022wkp,EventHorizonTelescope:2019dse} and the groundbreaking detection of gravitational waves by LIGO/Virgo/KAGRA \cite{LIGOScientific:2016aoc,KAGRA:2021vkt} have revolutionized the study of strong-field gravity, providing unprecedented opportunities to test general relativity and explore modified gravity theories. In \cite{Cardoso:2016rao}, the authors show that despite having a completely different quasinormal-mode spectrum, some wormholes exhibit a ringdown stage very similar to that of black holes, where the differences would only appear at late times. Other works demonstrate that, depending on the model parameters, the ringdown of wormholes can either completely mimic that of black holes at all times or exhibit consistent deviations in every phase \cite{Konoplya:2016hmd}. Thus, gravitational wave observations do not completely rule out the possibility of wormholes, giving us the opportunity to compare these objects with astrophysical observations in extreme regimes. 

Wormholes have captivated physicists as theoretical constructs that could bridge distant regions of spacetime or even connect entirely separate universes. First brought into the modern framework by Ellis \cite{Ellis:1973yv} and Bronnikov \cite{Bronnikov:1973fh}, and generalized by Morris and Thorne \cite{Morris1988WormholesIS}, traversable wormholes pose unique challenges to classical physics due to their reliance on exotic matter to sustain the throat. Such exotic matter typically violates key energy conditions, such as the null energy condition (NEC), raising questions about their physical plausibility. As a result, researchers have sought alternative formulations involving modified gravity theories or exotic field couplings to construct more realistic and physically acceptable wormhole solutions \cite{Muniz:2022eex,Mustafa:2023kqt,Nilton:2022hrp,Loewer:2024lai,Muniz:2024jzg,Battista:2024gud,DeFalco:2021ksd,Bambi:2021qfo,Magalhaes:2022esc,Jusufi:2020rpw,Magalhaes:2023har}. Some types of wormhole that have gained considerable attention in recent years are those that can be obtained in general relativity by considering a Dirac field. The Dirac field possesses the exotic properties necessary to generate these solutions without requiring additional exotic fields \cite{Konoplya:2021hsm,Kain:2023pvp,Kain:2023ann,Blazquez-Salcedo:2021udn,Blazquez-Salcedo:2020czn,Bolokhov:2021fil,Maldacena:2018gjk}.

In this context, Bronnikov \cite{Bronnikov:2000vy,Bronnikov:2022bud} demonstrated that nonlinear electrodynamics can effectively regularize black hole spacetimes while simultaneously supporting wormhole geometries. Other studies highlighted the role of nonlinear field interactions in generating the stress-energy tensor components required to stabilize wormhole throats \cite{Shaikh:2018yku,Javed:2022dfn,Canate:2022dzb}. Similarly, scalar fields, particularly phantom scalar fields characterized by negative kinetic energy, have been shown to naturally exhibit the exotic matter properties needed for compact objects and traversable wormholes \cite{Dzhunushaliev:2016xdt,Kamal:2018oxw}. The combination of nonlinear electrodynamics and scalar fields provides a versatile framework to explore diverse wormhole geometries within general relativity \cite{Crispim:2024dgd}.

The combination of phantom scalar fields with nonlinear electrodynamics can also generate solutions known as black bounces \cite{Bronnikov:2021uta,Rodrigues:2023vtm}. These spacetimes have a rich causal structure and can transition between regular black holes and wormholes \cite{Lobo:2020ffi}. The first black bounce solution was proposed by Simpson and Visser through a regularization process of the Schwarzschild metric \cite{Simpson:2018tsi}. This type of regularization can also be applied to other known solutions, considering, for example, rotating solutions \cite{Franzin:2021vnj,Lima:2023jtl}, solutions with cylindrical symmetry \cite{Lima:2022pvc,Bronnikov:2023aya,Lima:2023arg}, black holes in $2+1$ dimensions \cite{Furtado:2022tnb,Alencar:2024nxi}, or extra dimensions \cite{Crispim:2024yjz,Crispim:2024nou}. The electromagnetic sector of these solutions can be either magnetically or electrically charged \cite{Alencar:2024yvh}. In the scalar sector, different types of scalar fields can be used, such as $k$-essence models \cite{Pereira:2023lck,Pereira:2024gsl,Pereira:2024rtv}.

Recent advancements have expanded the theoretical landscape of wormhole physics. Rodrigues et al. \cite{Rodrigues:2022mdm} introduced the concept of black bounces with multiple throats and anti-throats, a topology characterized by extrema in the areal function of the metric. These extrema correspond to regions of enhanced curvature and add complexity to the wormhole's geometry. Such structures can be visualized using embedding diagrams, which highlight the intricate topologies associated with multiple throats. These developments underscore the importance of selecting appropriate metric functions and field couplings to model realistic wormholes. In \cite{Chew:2018vjp}, the authors find wormhole solutions in an alternative theory of gravity that can present multiple throats in the Jordan frame, but not in the Einstein frame. It is also possible to find this type of structure in general relativity by considering a phantom scalar field together with a Yang-Mills-Higgs field \cite{Chew:2020svi,Chew:2021vxh}.

Building on these foundational works, this study examines wormholes with multiple throats and anti-throats in the context of general relativity. By coupling nonlinear electrodynamics with phantom scalar fields, we explore how specific field configurations generate solutions consistent with Einstein's field equations. Inspired by approaches from Bazeia et al. \cite{Bazeia:2017rxo} and Vachaspati \cite{Vachaspati:2006zz}, the appropriate scalar field profiles  are employed to study the coupling functions and potentials associated with these geometries. These scalar field models enable the investigation of the physical properties of the wormhole, including its stability and the behavior of energy condition violations. 

A crucial aspect of this analysis involves exploring the energy conditions, following the methodologies outlined by Visser \cite{Visser:1995cc}. While energy condition violations are generally unavoidable in traversable wormholes, this study identifies parameter configurations that minimize such violations near the wormhole's throat. This represents an essential step toward the development of physically plausible wormhole models that may align with astrophysical observations. 

Thus, by integrating nonlinear electrodynamics and scalar fields into the study of wormholes with multiple throats, this work extends the theoretical framework of traversable wormholes. It demonstrates the flexibility in designing wormhole solutions with tailored geometric and physical properties, providing a robust foundation for future investigations into the exotic nature of spacetime.

The structure of this paper is as follows: Section II introduces the multi-throat wormhole solutions, focusing on their geometry and geodesic structure. Section III examines the field sources that generate these solutions, based on nonlinear electrodynamics and a phantom-like scalar field. In Section IV, we analyze the energy conditions associated with the wormhole configurations. Finally, Section V summarizes our findings and presents the concluding remarks.

\section{Wormhole spacetime}

The first and simplest wormhole spacetime was proposed independently by Ellis and Bronnikov \cite{Ellis:1973yv,Bronnikov:1973fh}, whose metric is given by
\begin{eqnarray}
  \label{ellis}  ds^2 =  dt^2 - dr^2 - (r^2 + d^2)d\Omega_2^2,
\end{eqnarray}
where $d\Omega_2^2 = d\theta^2 + \sin^2\theta d\varphi^2$ and  $d$, in this specific case, is the size of the throat, which in these coordinates is located at $r = 0$. With the metric written in this form \eqref{ellis}, $r$ is the proper radial distance and take values on $r \in (-\infty, +\infty)$ \cite{Morris1988WormholesIS}.

From this, several works with modifications or generalizations of this geometry have been carried out \cite{Kar:1995jz, Nilton:2022hrp, Bronnikov:2022bud,Canate:2022dzb,Crispim:2024dgd}.

On the other hand, in \cite{Rodrigues:2022mdm} the authors proposed a series of black bounce spacetimes with non-usual areas. These models can be studied in the context of wormholes by adapting the line element in such a way that it can be written as:
\begin{equation}
    ds^2=dt^2-dr^2-\Sigma^2(r)d\Omega_2^2.\label{line}
\end{equation}
The type of wormhole model is described by the function $\Sigma(r)$. The Kretschmann scalar of this metric is given by
\begin{eqnarray}
    \mathcal{K} = \frac{4 \left(2 \Sigma^2 \Sigma ''^2+\Sigma '^4-2 \Sigma '^2+1\right)}{\Sigma ^4},
\end{eqnarray}
where can we see that the geometry is everywhere regular if the following conditions are satisfied \cite{Lobo:2020ffi}:
\begin{itemize}
    \item $\Sigma(r)$ must be non-zero everywhere.
    \item $\Sigma'(r)$ and $\Sigma''(r)$ must be finite everywhere.
\end{itemize}

Among the black bounce models proposed in the original work, one that stands out is given by:
\begin{equation}
    \Sigma^2(r)=\left(d^2+r^2\right) e^{\frac{b^2}{c_3+r^2}},\label{model}
\end{equation}
where the constants $b, c_3$ and $d$ are parameters that influence the shape of the wormhole. This model was chosen because it generates regular solutions. To see this, it is enough to consider the following expressions for the first and second derivatives of the function $\Sigma$ of our model, given by
\begin{eqnarray}
    \Sigma'(r) &=& \frac{r \left[\left(c_3+r^2\right)^2-b^2 \left(d^2+r^2\right)\right]e^{\frac{b^2}{2(c_3+r^2)}}}{\left(c_3+r^2\right)^2  \left(d^2+r^2\right)^{3/2}},\\
   \Sigma''(r)&=& \frac{b^2 \left[-c_3^2 \left(d^2+3 r^2\right)+3 d^2 r^4+r^6\right] e^{\frac{b^2}{2(c_3+r^2)}}}{\left(c_3+r^2\right)^4
   \left(d^2+r^2\right)^{1/2}} +\frac{b^4 r^2 \sqrt{d^2+r^2
  }e^{\frac{b^2}{2(c_3+r^2)}}}{\left(c_3+r^2\right)^4}\\\nonumber
   &+&\frac{e^{\frac{b^2}{2(c_3+r^2)}} \left[2 b^2 c_3 r^2 \left(d^4-r^4\right)+c_3^4 d^2+4 c_3^3 d^2 r^2+6
   c_3^2 d^2 r^4+4 c_3 d^2 r^6+d^2 r^8\right]}{\left(c_3+r^2\right)^4 \left(d^2+r^2\right)^{3/2}}.
\end{eqnarray}

For $b = 0$, the conditions for regularity are trivially satisfied. In fact, $b = 0$ recovers the well-known Ellis-Bronnikov wormhole \cite{Ellis:1973yv,Bronnikov:1973fh}. Considering $b \neq 0$, if $c_3 = 0$, we see that the last conditions are no longer satisfied, since $\Sigma'$ and $\Sigma''$ will diverge and therefore the Kretschmann scalar would diverge. Therefore, the introduction of the parameters $b$ and $c_3$ aims to modify the classical geometry of the wormhole, allowing it to have multiple throats and anti-throats while ensuring that the geometry remains regular throughout the entire space.

Furthermore, as previously stated, depending on the choice of parameters, the shape of the wormhole can change significantly, exhibiting multiple throats/anti-throats.
It is important to note that in our model, $d$ no longer represents the size of the wormhole throat. The inclusion of the constants $b$ and $c_3$ affects not only the position of the throat but also the number of throats \cite{Rodrigues:2022mdm}.

In this chosen coordinate system, the area of a sphere at the radial coordinate $r$ is now defined as \cite{Lobo:2020ffi,Boonserm:2018orb}:
\begin{equation}
    A=4\pi \Sigma^2(r).
\end{equation}
In the study of wormholes, we can determine the presence of throats/anti-throats by analyzing the maxima and minima of the wormhole's area. This is widely discussed in the reference \cite{Rodrigues:2022mdm}, ensuring the presence of throats and anti-throats, which will be represented in the embedding diagrams below.

\subsection{Embedding Diagrams}

Embedding diagrams are commonly used as a way to facilitate the visualization of a curved surface \cite{Morris1988WormholesIS}. In our context, we are interested in embedding the two-dimensional spherically symmetric curved surface defined by the metric \eqref{line} with $t = constant$ and $\theta = \pi/2$
\begin{equation}
     ds^2=-dr^2-\Sigma^2(r) d\varphi^2,\label{line2}
\end{equation}
into a cylindrical Euclidean space defined by the line element
\begin{eqnarray}\label{linediagram}
    d\sigma^2 = -d\rho^2 - \rho^2d\varphi^2 - dz^2 = -\left[\left(\frac{d\rho(r)}{dr}\right)^2 + \left(\frac{dz(r)}{dr}\right)^2 \right]dr^2 - \rho^2d\varphi^2 .
\end{eqnarray}

Comparing the line elements \eqref{line2} and \eqref{linediagram} we get
\begin{eqnarray}
    \rho(r) = \Sigma (r),
\end{eqnarray}
\begin{equation}\label{zequacao}
    \frac{dz(r)}{dr} = \sqrt{1 - \Sigma'^2}.
\end{equation}

For consistency, we must have $z(r)$ well-defined throughout the all spacetime for all values of $r$ and $dz/dr \to 0$ for $r \to \pm \infty$, so we must have
\begin{equation}\label{consistencia}
    \Sigma'^2 \leq 1, \; \forall\; r \in (-\infty, +\infty),
\end{equation}
\begin{equation}
    \lim_{r \to \infty} \Sigma'^2 = 1.
\end{equation}

\begin{figure}[ht]
    \centering
    \includegraphics[width=1\linewidth]{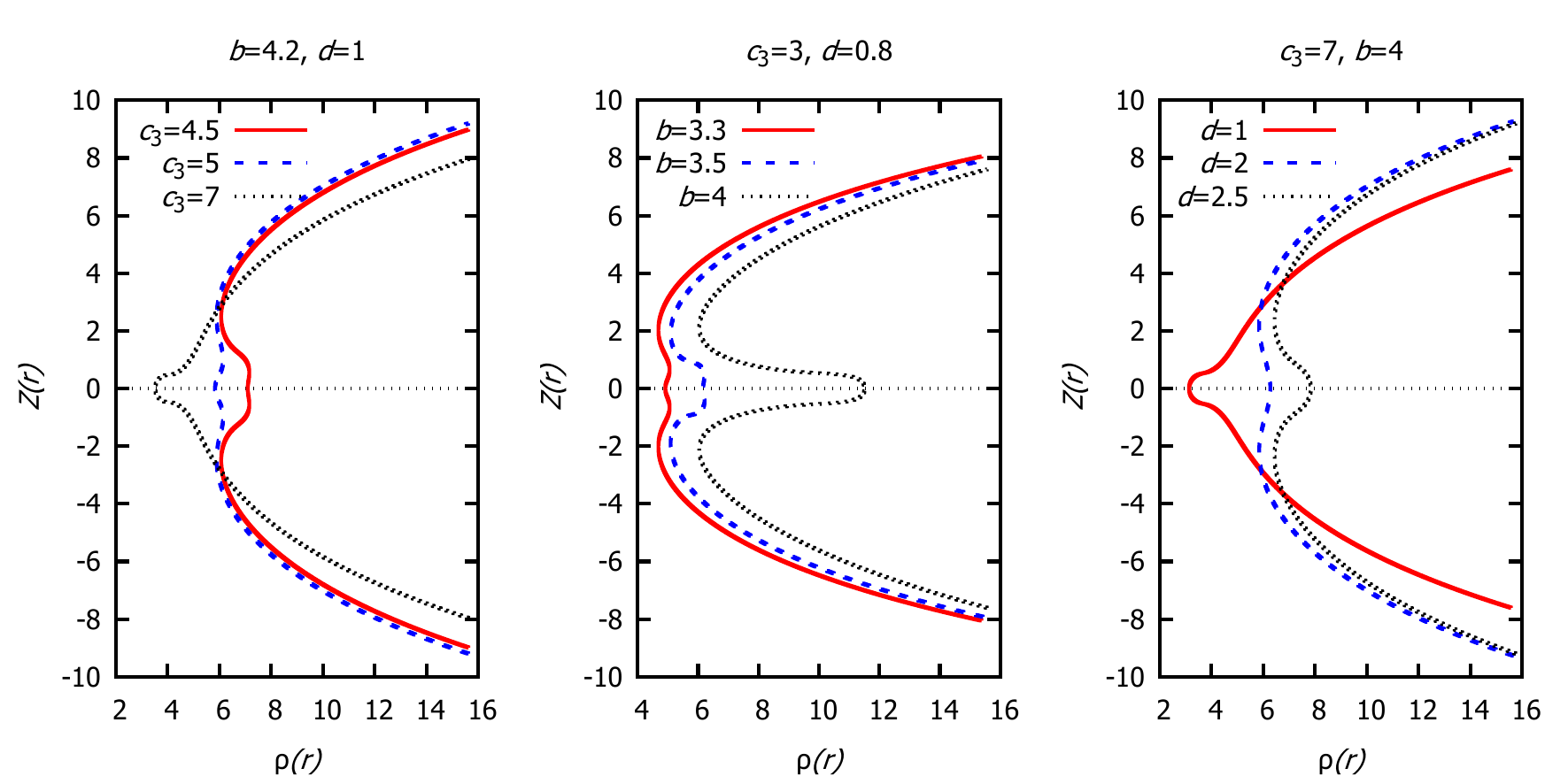}
    \caption{Parametric plot of $z(r)$ vs $\rho(r)$ showing the so-called isometric embedding diagrams for different choices of parameters. These curves stand out significantly compared to the typical catenary curve of the Ellis-Bronnikov wormhole.}
    \label{zvsp}
\end{figure}

 Let $\Sigma(r)$ be given by the equation \eqref{model}, it is straightforward to check that the above limit is satisfied for any values of the free parameters.  However, condition \eqref{consistencia} is not satisfied for all values of parameters $d$, $b$ and $c_3$, such that only certain combinations are allowed to have a consistent solution. Following the numerical integration of equation \eqref{zequacao} for different choice of parameters, we present in Fig.\ref{zvsp} the parametric plot of $z(r)$ vs $\rho(r)$ showing isometric embedding diagrams for our wormholes geometries characterized by the function $\Sigma(r)$ given in \eqref{model}. The fact that $\Sigma' \to \pm 1$ for $r \to \pm\infty$ shows that the geometry is asymptotically flat, which can also be visualized geometrically in the diagrams.

The plot in Fig. \ref{zvsp} is made with $\varphi$ fixed, so for a three-dimensional geometric visualization of how the two-dimensional curved surface given by \eqref{line2} appears when embedded in a three-dimensional Euclidean geometry, we should rotate the plot around the z-axis. By rotating these parametric plots around the $z(r)$-axis, we obtain the embedded surface diagrams of these wormholes, as presented in Fig. \ref{diagram}. It is important to note that in both the plots of Fig. \ref{zvsp} and the diagrams of Fig. \ref{diagram}, depending on the choice of parameters, multiple throats and anti-throats can be observed in the solution. Specifically, for $b = 4$, $c_3 = 3$, and $d = 0.8$, a pronounced anti-throat of the wormhole is evident. Furthermore, for smaller values of $b$, these throats/anti-throats become less pronounced, as for small values of $b$, the function $\Sigma (r)$ increasingly resembles the $\Sigma (r)$ function in the Ellis-Bronnikov case.

\begin{figure}[ht]
    \centering
    \includegraphics[width=0.95\linewidth]{diagramcqg.pdf}
    \caption{Embedded surface diagrams for different choices of parameters, where it is clearly possible to geometrically visualize the structure of multiple throats and anti-throats, differing significantly from the typical catenoid surface of the Ellis-Bronnikov wormhole. These diagrams were obtained by rotating the curves from Fig. \ref{zvsp} around the z-axis.}
    \label{diagram}
\end{figure}

Therefore, as demonstrated through the analysis of the minima and maxima of the area presented in \cite{Rodrigues:2022mdm}, and geometrically demonstrated here, our geometry describes a family of wormholes with multiple throats and anti-throats, depending on the choice of parameters $b, c_3$ and $d$. Next, we will briefly analyze the geodesic equations in this spacetime.

\subsection{Geodesics}
One of the ways to extract information about a spacetime is through the study of its geodesics. Thus, let us now analyze the trajectories in the spacetime described by the metric \eqref{line} and \eqref{model}, both for massive as massless particles. To do so, the complete set of geodesics equations are:
\begin{eqnarray}
\ddot{t}=0\ , \nonumber\\
\ddot{r}-\Sigma'(r)\Sigma(r)\dot{\theta}^2-\Sigma'(r)\Sigma(r)\sin^2\theta\dot{\varphi}^2=0\ ,\nonumber\\
\ddot{\theta}+2\frac{\Sigma'(r)}{\Sigma(r)}\dot{r}\dot{\theta}-\cos\theta\sin\theta\dot{\varphi}^2=0\ , \nonumber\\
\ddot{\varphi}+2\frac{\Sigma'(r)}{\Sigma(r)}\dot{r}\dot{\varphi}+2 \cot\theta\dot{\varphi}\dot{\theta}=0\ .
\label{geoeqs1}
\end{eqnarray}
Here, `$\cdot$' refer to derivatives with respect to an affine parameter of the trajectories. The first equation can be integrated directly to give $\dot{t}=E$, where $E$ is a constant that represents the total energy of the particle. Moreover, $\theta=\pi/2$ is a solution of the above set of equations, since the metric \eqref{line} is spherically symmetric. Then, the last equation in \eqref{geoeqs1} also provides a first integral that gives:
\begin{equation}
\Sigma^2\dot{\varphi}=\ell\ ,
\label{angle1}
\end{equation}
where $\ell$ is a constant. In addition, the line element  \eqref{line} provides the following constraint:
\begin{equation}
\dot{t}^2-\dot{r}^2-\Sigma^2(r)\dot{\varphi}^2= \epsilon,
\label{geoline1}
\end{equation}
where $\epsilon=1$ corresponds to massive particles whereas $\epsilon=0$ refers to massless ones. For each case, the equation for the radial coordinate $r$ yields:
\begin{equation}
\dot{r}^2=E^2-1-\frac{\ell^2}{\Sigma^2(r)}\ ,\quad \dot{r}^2=E^2-\frac{\ell^2}{\Sigma^2(r)}\ ,
\label{radialcoord1}
\end{equation}
where we have used the first integrals obtained above. The easiest trajectories to analyze are the radial ones, where $\ell=0$ above. For both cases, massive and massless, one gets:
\begin{equation}
\ddot{r}=0\ , \quad \dot{r}^2=E^2-\epsilon\ ,
\label{radialmotion}
\end{equation}
As one can easily point out, massive particles in the spacetime \eqref{line} just follow free paths radially, contrary to Schwarzschild spacetime for instance. For non-radial geodesics, it might be interesting to analyze the null trajectories, since they contribute to the appearance of the object when is observed by a far away observer. In this case, equation for massless particles in \eqref{radialcoord1} can be rewritten as:
\begin{equation}
 \dot{r}^2=\ell^2\left(\frac{1}{B^2}-V(r)\right)\ ,
 \label{radialcoord2}
 \end{equation}
where $B=\ell/E$ and $V(r)=\frac{1}{\Sigma^2(r)}$. One should note that there might be photon spheres, i.e. unstable circular orbits for photons as far as $V(r_c)=\frac{1}{B^2}$ and $V'(r_c)=0$.

\begin{figure}
    \centering
    \includegraphics[width=1\linewidth]{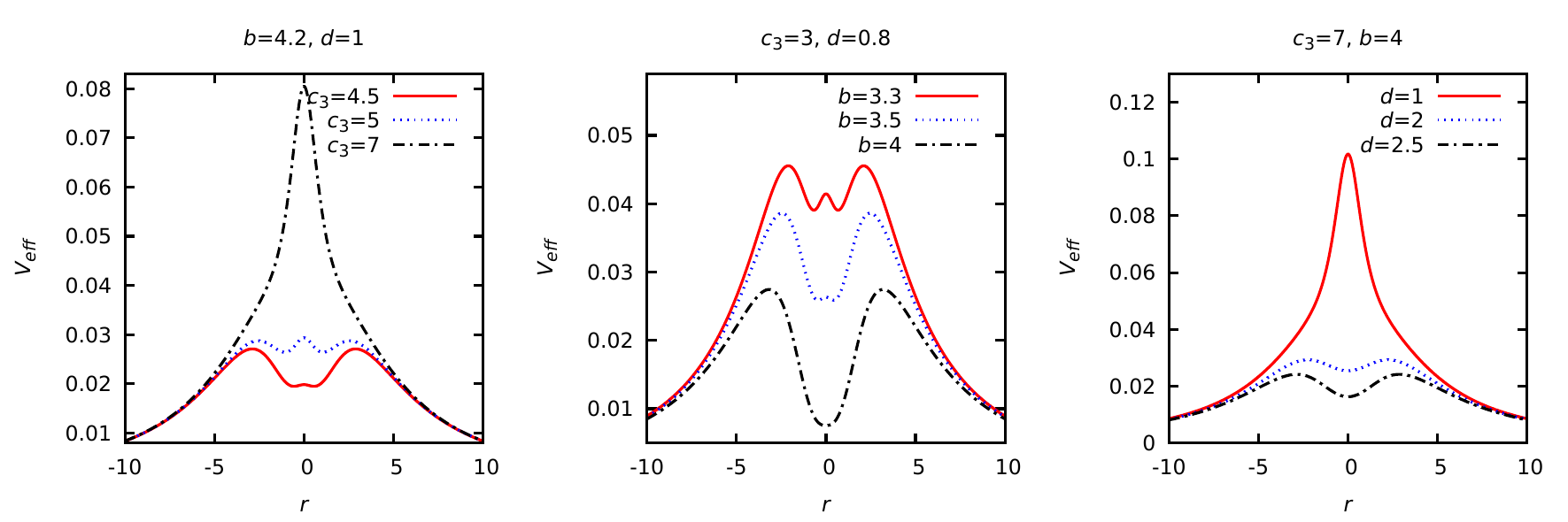}
    \caption{Effective potential for null geodesics on the background of the wormhole solution \eqref{model}, as a function of the radial coordinate for different values of the parameters $b$, $d$, and $c_3$.}
    \label{fig:Veffective}
\end{figure}

In Fig. \ref{fig:Veffective}, we show the shape of the effective potential for massless particles in the presence of the wormhole, considering different parameter values. We can observe that, depending on the choice of parameters, the potential may exhibit several maxima and minima. This implies that there can be multiple orbits, both unstable and stable, for photons. The maxima of the effective potential correspond to the locations of the wormhole throats (i.e., minima in the wormhole area), while the minima of the effective potential correspond to the locations of the wormhole anti-throats (i.e., maxima in the area). The fact that each throat has an unstable orbit for photons is consistent with the result presented in \cite{Xavier:2024iwr}.

Moreover, one can obtain a more proper equation for analyzing this type of trajectories by dividing \eqref{angle1} and \eqref{radialcoord2}, leading to:
\begin{equation}\label{angleradioeq}
\frac{d\varphi}{dr}=\pm \frac{B}{\Sigma^2\sqrt{1-\frac{B^2}{\Sigma^2}}}\ .
 \end{equation}
This equation can be used to analyze the null trajectories around the object, which can provide the way the object looks for an external observer when illuminated by an accretion disk.

Therefore, in summary, we have seen how the introduction of the function \eqref{model} significantly changes the spacetime structure of the wormhole. In the next section, we will show how this geometry can emerge as a solution from general relativity when considering the coupling of a partially phantom field with a nonlinear electrodynamics.

\section{Field Sources}
Usually, wormhole solutions can be obtained by considering a scalar field theory with a negative kinetic term in the Lagrangian. However, in our case, we find that the components of the Einstein tensor, $G_{\mu\nu}$, are given by
\begin{eqnarray}
    {G^0}_0&=&-\frac{2 \Sigma''}{\Sigma}-\frac{\Sigma'^2}{\Sigma^2}+\frac{1}{\Sigma^2},\\
    {G^1}_1&=&\frac{1}{\Sigma^2}-\frac{\Sigma'^2}{\Sigma^2},\\
    {G^2}_2&=&{G^3}_3=-\frac{\Sigma''}{\Sigma}.
\end{eqnarray}
From the form of the components of the Einstein tensor, in general, we see that a scalar field, even with a potential, would not be sufficient to act as a source since, for a scalar field, we have the relation ${T^0}_{0} = {T^2}_{2}$, on the components of the stress-energy tensor. Therefore, additional fields are required to generate this solution, such as a nonlinear electromagnetic field. It is important to emphasize that, for specific cases of $\Sigma$, it is possible that ${G^0}_{0}={G^2}_{2}$.

We will now verify what type of source fields can generate this type of wormhole. To this end, we will check if the presence of a nonlinear electromagnetic field together with a scalar field can generate this solution in general relativity. This theory is described by the action:
\begin{equation}
    S=\int \sqrt{\left|g\right|}d^4x\left[R - 2 h\left(\phi\right) g^{\mu\nu}\partial_\mu \phi\partial_\nu \phi
   		+2V(\phi) + L(F)\right],    			\label{Action}
\end{equation}
where $\phi$ is the scalar field, $V(\phi)$ is its potential, $L(F)$ is the nonlinear electrodynamics Lagrangian, $F=F^{\mu\nu}F_{\mu\nu}$, and $F_{\mu\nu} = \partial_\mu A_\nu - \partial_\nu A_\mu$ is the electromagnetic field tensor. The function $h\left(\phi\right)$ will determine if the scalar field is phantom, $h\left(\phi\right)<0$, or standard, $h\left(\phi\right)>0$. This type of action is widely used in the literature to determine the field sources of black bounce solutions and even wormholes \cite{Rodrigues:2025plw,Alencar:2024nxi,Alencar:2024yvh,Bronnikov:2023aya,Rodrigues:2023vtm,Bolokhov:2024sdy,Bronnikov:2022bud,Bronnikov:2021uta,Bronnikov:2016xvj}.

From the variation of action \eqref{Action} with respect to $A_\mu$, $\phi$, and $g^{\mu\nu}$, we obtain the equations of motion for the electromagnetic, scalar, and gravitational fields, which are given, respectively, by:
\begin{eqnarray}
    &&\nabla_\mu \left[L_F F^{\mu\nu}\right] =\frac{1}{\sqrt{\left|g\right|}}\partial_\mu \left[\sqrt{\left|g\right|}L_F F^{\mu\nu}\right]=0,\label{eq-F}\\
     &&2h\left(\phi\right) \nabla_\mu \nabla^\mu\phi +\frac{dh\left(\phi\right)}{d\phi}\partial^\mu\phi\partial_\mu\phi=-  \frac{dV(\phi)}{d\phi} ,     \label{eq-phi}\\  
      &&G_{\mu\nu} =R_{\mu\nu}-\frac{1}{2}R g_{\mu\nu} =T[\phi]_{\mu\nu}  + T[F]_{\mu\nu},				\label{eq-Ein}
\end{eqnarray}
where $L_F=d L/d F$, $T[\phi]_{\mu\nu}$ and $T[F]_{\mu\nu}$ are the stress-energy tensors of the scalar and electromagnetic fields, respectively,
\begin{eqnarray}                
   T[F]_{\mu\nu} = \frac{1}{2} g_{\mu\nu} L(F ) - 2L_F {F_{\nu}}^{\alpha} F_{\mu \alpha},\label{SETF}\\
    T[\phi]_{\mu\nu} =  2 h\left(\phi\right) \partial_\nu\phi\partial_\mu\phi 
   - g_{\mu\nu} \big (h\left(\phi\right) \partial^\alpha \phi \partial_\alpha \phi - V(\phi)\big).\label{SETphi}
\end{eqnarray}

Solving the Maxwell equations for a magnetically charged, static, and spherically symmetric object, we obtain that the magnetic field is given by:
\begin{equation}
    F_{23}=q \sin \theta,
\end{equation}
and the scalar $F$ is
\begin{equation}
    F=\frac{2q^2}{ \Sigma^2(r)},
\end{equation}
where $q$ is the magnetic charge.

Considering the line element given by \eqref{line}, the field equations are given by:
\begin{eqnarray}
    -\frac{L}{2}-\frac{2 \Sigma''}{\Sigma}-\frac{\Sigma'^2}{\Sigma^2}+\frac{1}{\Sigma^2}-V-h  \phi '^2=0,\label{eqmov1}\\
  -\frac{L}{2}+\frac{1-\Sigma'^2}{\Sigma^2}-V+h \phi '^2=0,\label{eqmov2}\\
   -\frac{L}{2}+\frac{2 q^2 L_F}{\Sigma^4}-\frac{\Sigma''+\Sigma \left(V+h \phi '^2\right)}{\Sigma}=0,\label{eqmov3}\\
   -\frac{4  h \Sigma' \phi '}{\Sigma}+\frac{V'}{\phi '}-\phi ' h' -2 h \phi ''=0.\label{eqmovphi}
\end{eqnarray}

By combining equations \eqref{eqmov2} and \eqref{eqmov3}, we can write the functions related to the electromagnetic field as:
\begin{eqnarray}
    L&=&-\frac{2 \left(\Sigma'^2+\Sigma^2 \left(V-h  \phi '^2\right)-1\right)}{\Sigma^2},\label{L-geral}\\
   L_F&=&\frac{\Sigma^2 \left(\Sigma \Sigma''-\Sigma'^2+2 \Sigma^2 h \phi '^2+1\right)}{2 q^2}.\label{LF-geral}
\end{eqnarray}
As we can note, for the electromagnetic functions to be determined, we first need to obtain the functions related to the scalar field.

From the equations of motion, we can also write:
\begin{equation}
   h(\phi)\phi '(r)^2= -\frac{\Sigma''}{\Sigma}.\label{eqhphi}
\end{equation}
Depending on the wormhole model, the function $\Sigma(r)$ is simple enough for $h=\pm1$, so that the scalar field can be obtained analytically from the equation above. For our model, this becomes more complicated, so we will follow the procedure used by \cite{Bronnikov:2022bud,Bolokhov:2024sdy,Alencar:2024nxi}. Through this method, we impose the scalar field as a smooth function and use relation \eqref{eqhphi} to obtain the function $h(\phi)$.

Written
\begin{equation}
    h(\phi)=-\frac{\Sigma''}{\Sigma \phi '^2},\label{hphigen}
\end{equation}
we can simplify the relations \eqref{L-geral} and \eqref{LF-geral}, which are given by
\begin{equation}
    L=-\frac{2 \left(\Sigma \Sigma''+\Sigma'^2+\Sigma^2 V-1\right)}{\Sigma^2},\quad \mbox{and} \quad
    L_F=-\frac{\Sigma^2 \left(\Sigma \Sigma''+\Sigma'^2-1\right)}{2 q^2}.
\end{equation}
From \eqref{eq-phi}, we obtain
\begin{equation}
    V'=\frac{-\Sigma \Sigma'''-3 \Sigma' \Sigma''}{\Sigma^2}.\label{v_gen}
\end{equation}
In this way, once we know the form of the function $\Sigma$, we can obtain the functions $L$, $L_F$, and $V$, regardless of the scalar field model we choose. Now, let us examine some scalar field models.

\subsection{Model $\phi=\arctan(r/d)$}
We will impose the scalar field as the monotonic function $\phi=\arctan(r/d)$. This model is considered in several works \cite{Bronnikov:2022bud,Bolokhov:2024sdy,Alencar:2024nxi}. Using the relation \eqref{hphigen}, we obtain the function $h(\phi)$, which is given by:
\begin{equation}
    h(\phi(r))=-\frac{\left(d^2+r^2\right)^2 \Sigma''(r)}{d^2 \Sigma(r)}.\label{hphi}
\end{equation}

\begin{figure}
    \centering
    \includegraphics[width=1.0\linewidth]{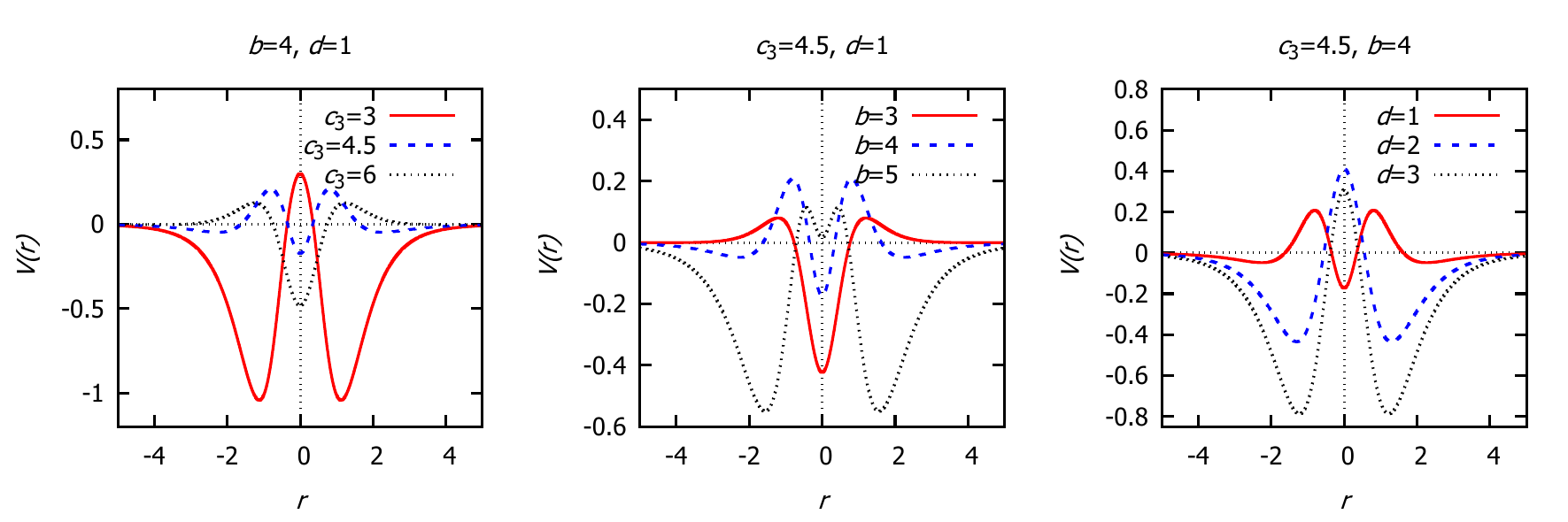}
    \includegraphics[width=1.0\linewidth]{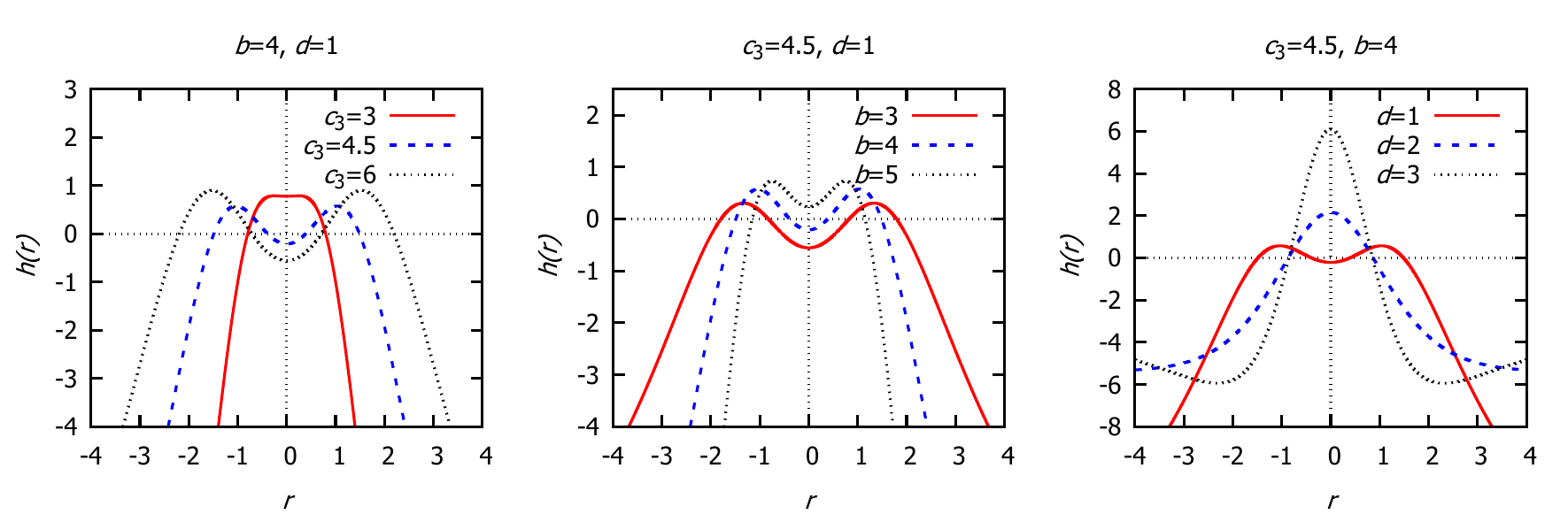}
    \caption{Graphical behavior  of the functions $V(\phi(r))$ (top panel) and $h(\phi(r))$ (bottom panel) in terms of the coordinate $r$ for different combinations of the wormhole parameters $d, b$ and $c_3$ and the model $\phi = \arctan(r/d)$.} 
    \label{fig:V-h-r}
\end{figure}

\begin{figure}
    \centering
    \includegraphics[width=1.0\linewidth]{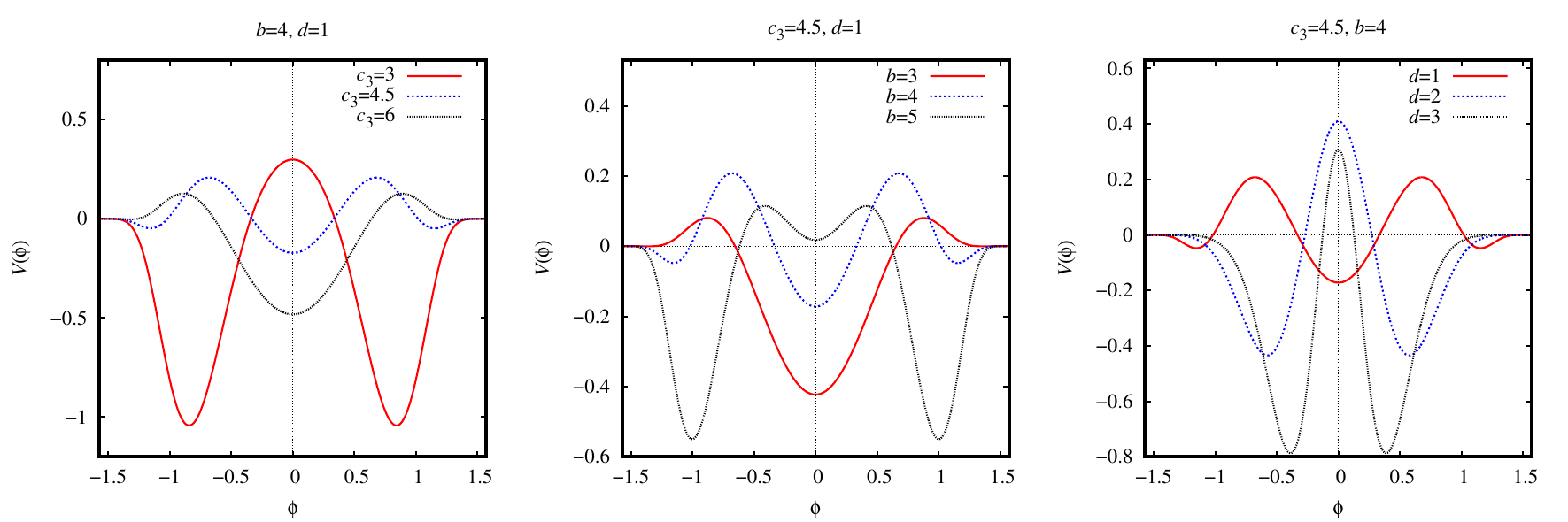}
    \includegraphics[width=1.0\linewidth]{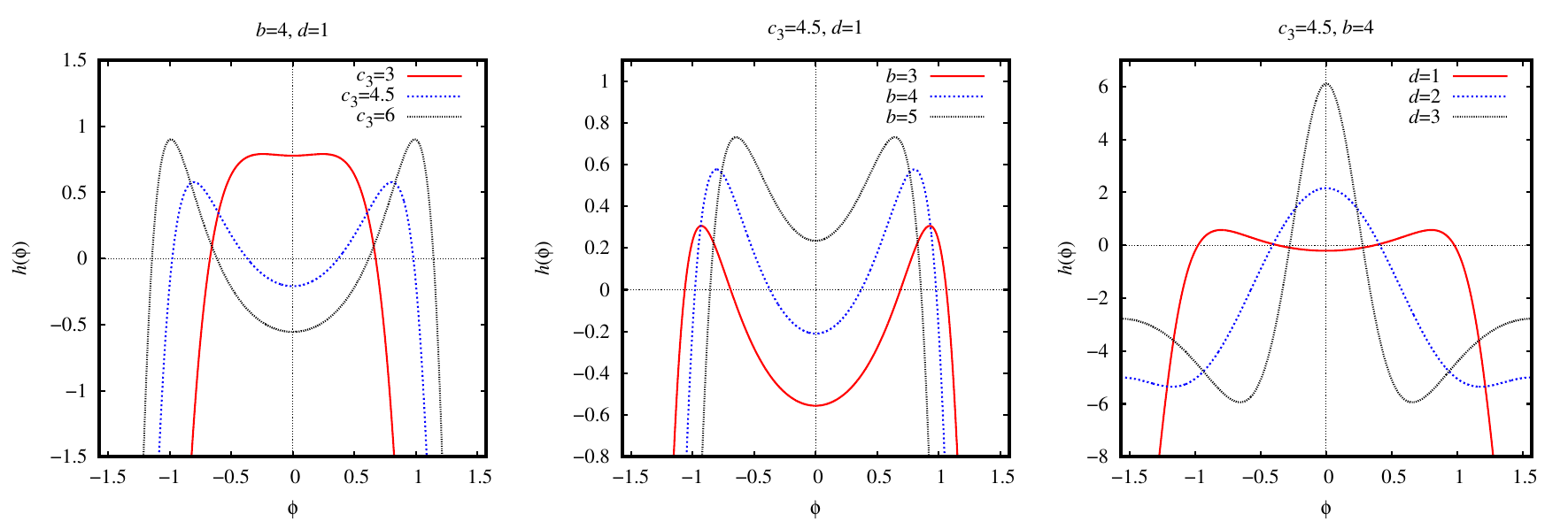}
    \caption{Graphical behavior of the functions $V(\phi)$ (top panel) and $h(\phi)$ (bottom panel) in terms of the scalar field $\phi$ for different combinations of the wormhole parameters $b, d$ and $c_3$.}
    \label{fig:V-h-phi}
\end{figure}

As we have the form of $\Sigma(r)$, we can integrate \eqref{v_gen}, which results in
\begin{eqnarray}
    V &=& 2 b^2 \left[\frac{b^4 c_3}{5 \left(c_3+r^2\right)^5}-\frac{b^4-6 b^2 c_3}{4
   \left(c_3+r^2\right)^4}+\frac{4 c_3-\frac{b^2 \left(c_3-3 d^2\right)}{c_3-d^2}}{2 \left(c_3+r^2\right)^3}+\frac{3 \left(b^2 d^2+c_3^2-d^4\right)}{\left(c_3-d^2\right)^3 \left(c_3+r^2\right)}\right.\nonumber\\
   &+&\left.\frac{3 \left(b^2 d^2+c_3^2-d^4\right)}{2
   \left(c_3-d^2\right)^2 \left(c_3+r^2\right)^2}+\frac{3 \left(b^2 d^2+c_3^2-d^4\right)
   }{\left(c_3-d^2\right)^4}\ln \left(\frac{d^2+r^2}{c_3+r^2}\right)\right].
\end{eqnarray}
Explicitly, in terms of the radial coordinate, the function  $h(\phi(r))$ can be written as:
\begin{eqnarray}
    h(\phi(r))=-\frac{1}{d^2 \left(c_3+r^2\right)^4}\left\{b^4 r^2 \left(d^2+r^2\right)^2-c_3^2 \left[b^2 \left(d^2+r^2\right) \left(d^2+3 r^2\right)-6 d^2 r^4\right]\right.\nonumber\\
    \left.+2 b^2 c_3 r^2 \left(d^4-r^4\right)+b^2 r^4 \left(d^2+r^2\right) \left(3
   d^2+r^2\right)+c_3^4 d^2+4 c_3^3 d^2 r^2+4 c_3 d^2 r^6+d^2 r^8\right\}.\label{h-r-model1}
\end{eqnarray}
The behavior of the functions $V(\phi(r))$ and $h(\phi(r))$ is shown in Fig. \ref{fig:V-h-r}. We observe that the functions are well-behaved, and the sign of  $h(r)$ changes several times with the radial coordinate. For more distant points, we notice that the scalar field behaves as a phantom field. Depending on the choice of parameters, the scalar field in the central region can either be of the phantom type or standard.
The function  $\phi(r)$ can be inverted so that we can obtain the form of the functions $h(\phi)$ and $V(\phi)$, which are written as:
\begin{eqnarray}
    V(\phi)&=&\frac{b^2}{10}  \left[\frac{4 b^4 c_3}{\left(c_3+d^2 \tan ^2\phi\right)^5}-\frac{5 b^2 \left(b^2-6 c_3\right)}{\left(c_3+d^2 \tan ^2\phi\right)^4}+\frac{40 c_3
   \left(c_3-d^2\right)-10 b^2 \left(c_3-3 d^2\right)}{\left(c_3-d^2\right) \left(c_3+d^2 \tan ^2\phi\right)^3}\right.\nonumber\\
   &+&\frac{60 \left(d^2 (b-d)
   (b+d)+c_3^2\right)}{\left(c_3-d^2\right)^3 \left(c_3+d^2 \tan ^2\phi\right)}+\frac{30 \left(d^2 (b-d) (b+d)+c_3^2\right)}{\left(c_3-d^2\right)^2 \left(c_3+d^2 \tan
   ^2\phi\right)^2}\nonumber\\
   &+&\left.\frac{60 \left(d^2 (b-d) (b+d)+c_3^2\right) }{\left(c_3-d^2\right)^4}\ln \left(\frac{d^2 \sec ^2\phi}{c_3+d^2 \tan ^2\phi
  }\right) \right],\label{v-phi-model1}\\
  h(\phi)&=&-\frac{1}{\left(c_3+d^2 \tan ^2\phi\right)^4}\left\{d^2 \tan ^2\phi \left[b^4 d^2+4 c_3^3+2 b^2 c_3 \left(d^2-2 c_3\right)+d^4 \left(b^2+d^2\right) \tan ^6\phi\right.\right.\nonumber\\
  &+&\left.\tan ^2\phi \left(2 b^4 d^2+3 b^2
   \left(d^4-c_3^2\right)+6 c_3^2 d^2\right)+d^2 \tan ^4\phi \left(b^4+4 d^2 \left(b^2+c_3\right)-2 b^2 c_3\right)\right]\nonumber\\
   &+&\left.c_3^2 (c_3-b d) (b
   d+c_3)\right\}.\label{h-phi-model1}
\end{eqnarray}
The form of the functions is somewhat complicated and not very clear. Therefore, it is more effective to analyze the behavior of the functions graphically. From Fig. \ref{fig:V-h-phi}, we see that both functions are symmetric under the transformation $\phi \rightarrow -\phi$. The scalar field tends to a constant for large values of the radial coordinate, $\phi(x \rightarrow \pm \infty) \rightarrow \pm \pi/2$, and in this limit, $h(\phi)$ tends to a constant while $V(\phi)$ tends to zero.

\begin{figure}
    \centering
    \includegraphics[width=1.0\linewidth]{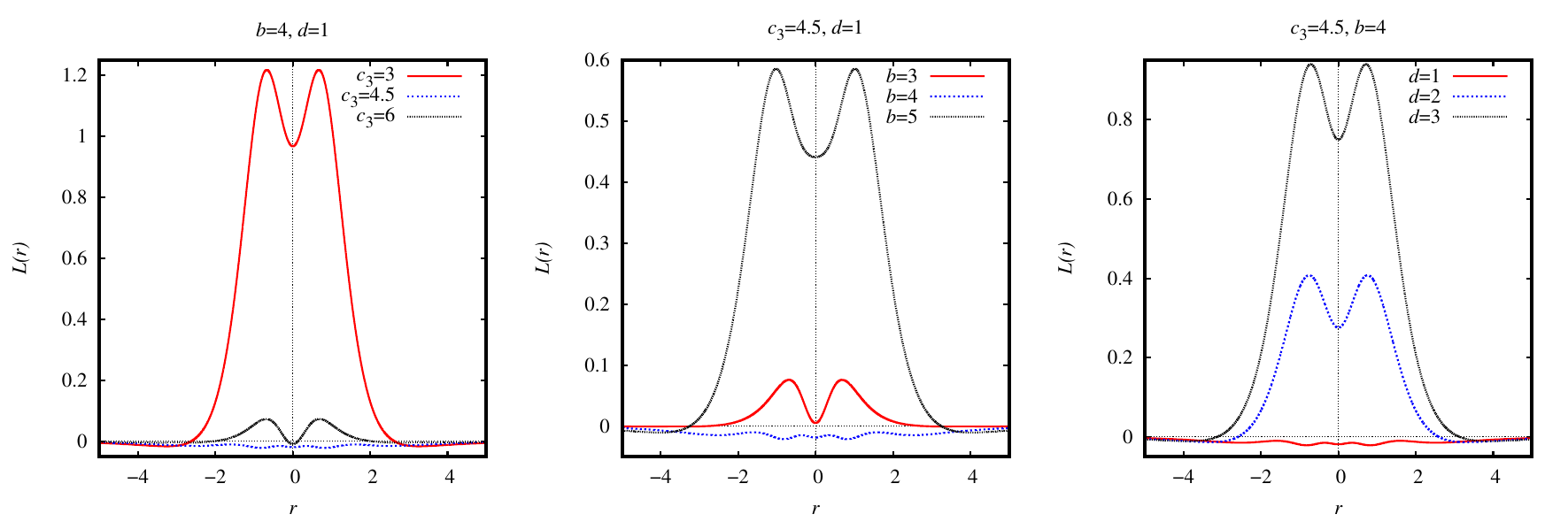}
    \includegraphics[width=1.0\linewidth]{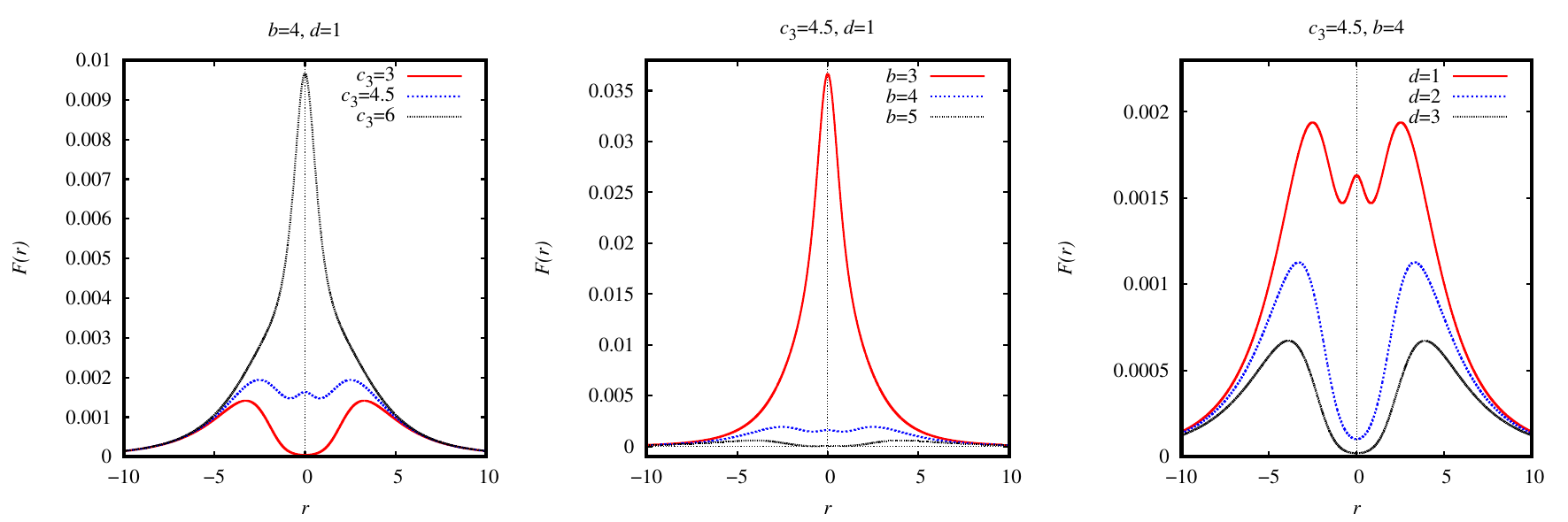}
    \caption{Behavior of the electromagnetic Lagrangian $L(r)$ (top panel) and electromagnetic scalar invariant $F(r)$ (bottom panel) in terms of the coordinate $r$ for different combinations of the wormhole parameters $b, d$ and $c_3$.}
    \label{fig:L-F-r}
\end{figure}

\begin{figure}
    \centering
    \includegraphics[width=1.0\linewidth]{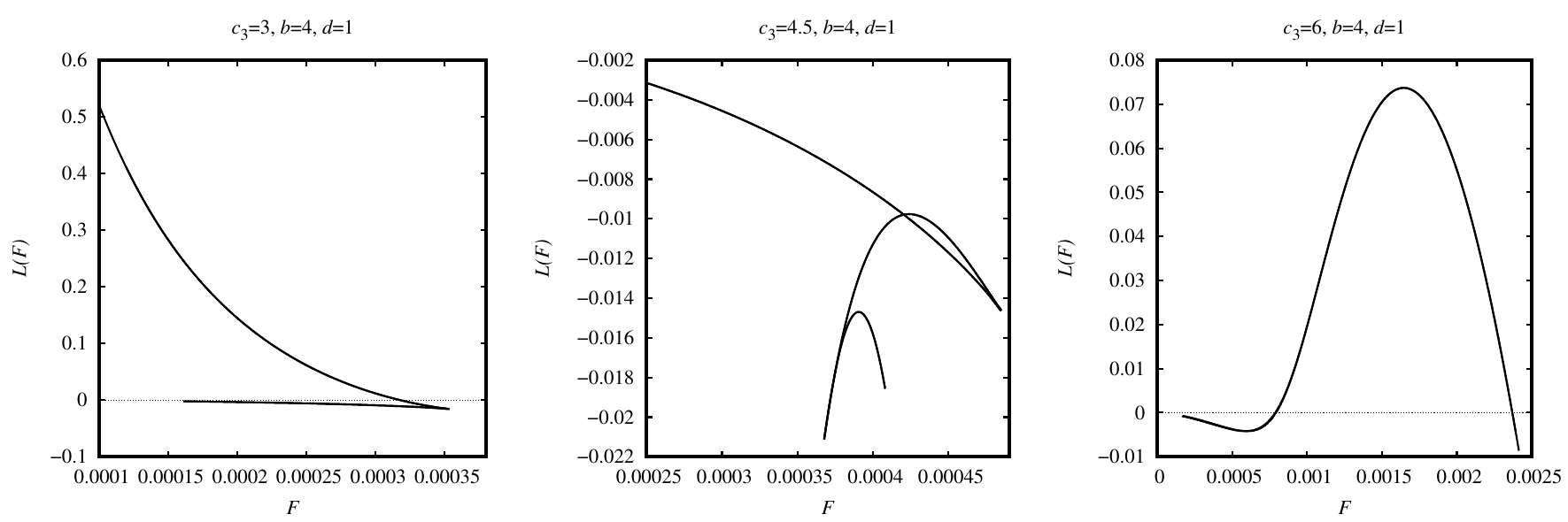}
    \caption{Behavior of the electromagnetic Lagrangian $L(F)$ in terms of the scalar $F$ for different combinations of the wormhole parameters $b, d$ and $c_3$.}
    \label{fig:L-F}
\end{figure}

Now that we have the quantities related to the scalar field, the functions related to the electromagnetic field are given by:
\begin{eqnarray}\label{L_r}
    L&=&-\frac{2 b^4 r^2}{\left(c_3+r^2\right)^4}-\frac{2 r^2 \left[\left(c_3+r^2\right)^2-b^2 \left(d^2+r^2\right)\right]^2}{\left(c_3+r^2\right)^4 \left(d^2+r^2\right)^2}+\frac{2
   e^{-\frac{b^2}{c_3+r^2}}}{d^2+r^2}-\frac{2
   d^2}{\left(d^2+r^2\right)^2}\nonumber\\
   &+&\frac{2 b^2 \left[3 r^2 \left(c_3-d^2\right)+c_3 d^2-r^4\right]}{\left(c_3+r^2\right)^3 \left(d^2+r^2\right)}-4 b^2 \left\{-\frac{3 \left[d^2 (b-d) (b+d)+c_3^2\right] }{\left(c_3-d^2\right)^4}\ln \left(\frac{c_3+r^2}{d^2+r^2}\right)\right.\nonumber\\
   &-&\frac{1}{20
   \left(c_3+r^2\right)^5}\left[\frac{10
   \left(c_3+r^2\right)^2 \left(b^2 \left(c_3-3 d^2\right)+4 c_3 \left(d^2-c_3\right)\right)}{c_3-d^2}+5 \left(b^4-6 b^2 c_3\right) \left(c_3+r^2\right)\right.\nonumber\\
   &-&\left.\left.\frac{60
   \left(c_3+r^2\right)^4 \left(d^2 (b^2-d^2) +c_3^2\right)}{\left(c_3-d^2\right)^3}-\frac{30 \left(c_3+r^2\right)^3 \left(d^2 (b^2-d^2)+c_3^2\right)}{\left(c_3-d^2\right)^2}-4 b^4 c_3\right]\right\},
\end{eqnarray}
\begin{eqnarray}
      L_F&=&\frac{\left(d^2+r^2\right) e^{\frac{b^2}{c_3+r^2}}}{2 q^2
   \left(c_3+r^2\right)^4} \left\{\left(c_3+r^2\right)^4-e^{\frac{b^2}{c_3+r^2}} \left[r^2 \left(c_3^2 \left(4 c_3-5 b^2\right)+2 b^2 d^2
    \left(b^2+c_3\right)\right)\right.\right.\nonumber\\
   &-&\left.\left.r^6 \left(b^2-4 c_3\right)+r^4 \left(2 b^4+3 b^2 \left(d^2-2 c_3\right)+6 c_3^2\right)+c_3^2 (c_3-b d) (b d+c_3)+r^8\right]\right\},\\
    F&=&\frac{2 q^2 e^{-\frac{2 b^2}{c_3+r^2}}}{\left(d^2+r^2\right)^2}.
\end{eqnarray}
 Although quite complex, the electromagnetic functions obey the consistency relation:
\begin{equation}
    L_F \frac{dF}{dr}-\frac{dL}{dr}=0.
\end{equation}
This consistency relation shows that, even though they are obtained independently through Maxwell and Einstein equations, the functions related to the electromagnetic field, $F$, $L$, and $L_F$, are interrelated. In Fig. \ref{fig:L-F-r}, we see the behavior of the functions $F$ and $L$ in terms of the radial coordinate, and we notice that these functions are well-behaved and symmetric under the transformation $r \rightarrow -r$. Something we can observe is that, depending on the chosen parameter values, the function $F(r)$ may present several maxima and minima, and we cannot analytically invert $r(F)$ to explicitly express the function $L(F)$. For each non-zero maximum/minimum of the function $F(r)$, there will be a cusp present in the function $L(F)$ \cite{Bronnikov:2000vy,Rodrigues:2020pem}. The presence of these cusps becomes evident in Fig. \ref{fig:L-F}. Depending on the parameter values, there may be cusps or we may have only a smooth function.

In this way, the chosen model for the scalar field is consistent with the method and is able to satisfy the field equations. Thus, the behavior of all functions related to the source fields is described. Next, we will consider another scalar field model.

\subsection{Model $\phi= \tanh(r/d)$}

In the previous subsection, we considered a scalar field model given by the arctangent function. Now, we explore an alternative choice to investigate how different scalar field profiles affect the structure of the functions, such as the potential $V(\phi)$. Specifically, we will consider $\phi = \tanh (r/d)$, a form that naturally arises in studies of kinks for potentials of the type $V(\phi) = (\phi^2 - 1)^2/2$ \cite{Vachaspati:2006zz,Bazeia:2017rxo}. However, we focus solely on the scalar field profile itself, without considering the form of the potential.

\begin{figure}
    \centering
    \includegraphics[width=1.0\linewidth]{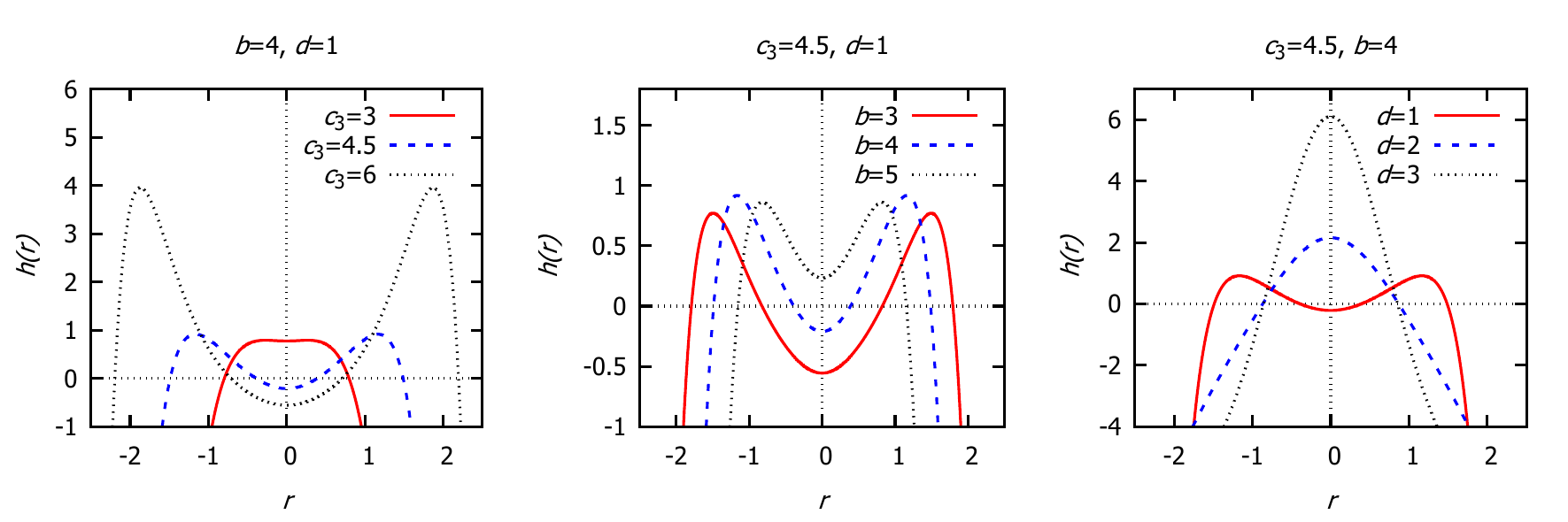}
    \caption{Behavior of the function $h(\phi(r))$ in terms of the coordinate $r$ for different combinations of the wormhole parameters $b, d$ and $c_3$ and the model $\phi=\tanh{(r/d)}$.}
    \label{fig:V-h-r_model2}
\end{figure}

Since the forms of the functions $L(r)$, $L_F(r)$, and $V(r)$ do not depend on the chosen scalar field, they remain the same as in the previous model. However, the forms of $h(r)$, $h(\phi)$, and $V(\phi)$ are modified. From \eqref{hphigen}, we find:
\begin{eqnarray}
    h(\phi(r))&=&-\frac{d^2 \cosh ^4\left(\frac{r}{d}\right) }{\left(c_3+r^2\right)^4
   \left(d^2+r^2\right)^2}\left[r^8 \left(b^2+d^2\right)+d^2 r^2 \left(b^4 d^2+2 b^2 c_3 \left(d^2-2 c_3\right)+4 c_3^3\right)\right.\nonumber\\
   &+&r^4 \left(2 b^4 d^2+3 b^2
   \left(d^4-c_3^2\right)+6 c_3^2 d^2\right)+r^6 \left(b^4+4 d^2 \left(b^2+c_3\right)-2 b^2 c_3\right)\nonumber\\
   &+&\left.c_3^2 d^2 (c_3-b d) (b d+c_3)\right].\label{h-r-model2}
\end{eqnarray}
As we can see, \eqref{h-r-model1} and \eqref{h-r-model2} are different. In Fig. \ref{fig:V-h-r_model2}, we analyze the behavior of $h(\phi(r))$. We can observe that, despite the functions \eqref{h-r-model1} and \eqref{h-r-model2} being different, both exhibit negative values in the same regions. This indicates that both scalar field choices have the same phantom-like profile.

\begin{figure}
    \centering
    \includegraphics[width=1.0\linewidth]{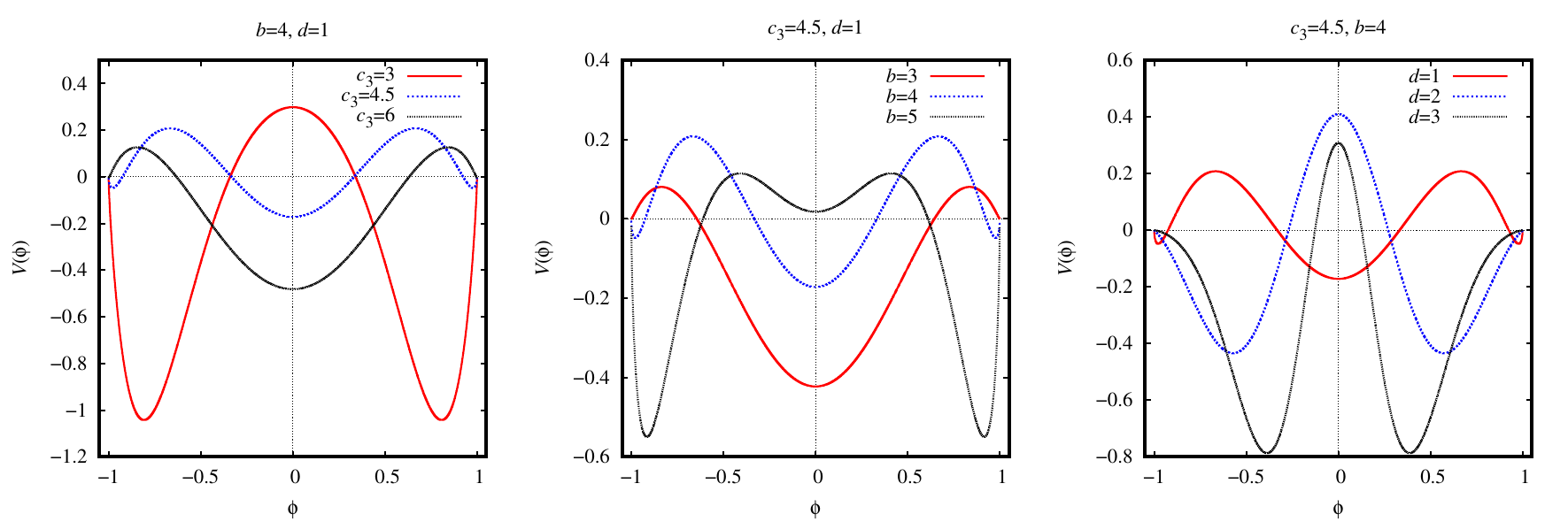}
    \includegraphics[width=1.0\linewidth]{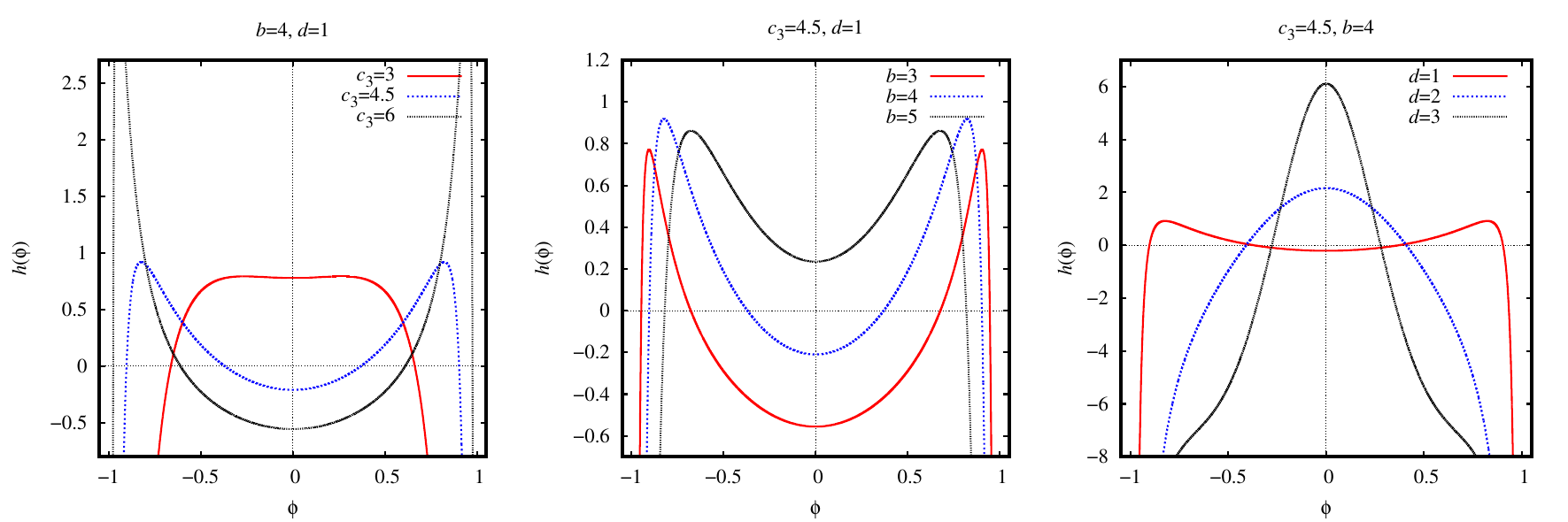}
    \caption{Graphical behavior of the functions $V(\phi)$ (top panel) and $h(\phi)$ (bottom panel) in terms of the scalar field $\phi$ for different combinations of the wormhole parameters $b, d$ and $c_3$.}
    \label{fig:V-h-phi_model2}
\end{figure}

Even though $V(r)$ does not change its form for this second model, $V(\phi)$ is different since the scalar field has changed. The analytical forms of $V(\phi)$ and $h(\phi)$ are given by:
\begin{eqnarray}
    V(\phi)&=&2 b^2 \left\{\frac{3 \left(b^2 d^2+c_3^2-d^4\right) }{\left(c_3-d^2\right)^4}\ln \left[\frac{d^2 \left(\tanh ^{-1}(\phi )^2+1\right)}{c_3+d^2 \tanh ^{-1}(\phi
   )^2}\right] \right.\nonumber\\
   &&+\frac{1}{20 \left(c_3+d^2 \tanh ^{-1}(\phi )^2\right)^5}\left[4 b^4 c_3+\frac{60 \left(b^2 d^2+c_3^2-d^4\right) \left(c_3+d^2 \tanh ^{-1}(\phi )^2\right)^4}{\left(c_3-d^2\right)^3}\right.\nonumber\\
   &+&\frac{30
   \left(b^2 d^2+c_3^2-d^4\right) \left(c_3+d^2 \tanh ^{-1}(\phi )^2\right)^3}{\left(c_3-d^2\right)^2}-5 \left(b^4-6 b^2 c_3\right) \left(c_3+d^2 \tanh ^{-1}(\phi )^2\right)\nonumber\\
   &-&\left.\left.\frac{10 \left(b^2 \left(c_3-3 d^2\right)+4 c_3 \left(d^2-c_3\right)\right)
   \left(c_3+d^2 \tanh ^{-1}(\phi )^2\right)^2}{c_3-d^2}\right]\right\},
\end{eqnarray}
\begin{eqnarray}
   h(\phi)&=&-\frac{\left(c_3+d^2 \tanh ^{-1}(\phi )^2\right)^{-4}}{\left(\phi ^2-1\right)^2
   \left(\tanh ^{-1}(\phi )^2+1\right)^2 }\left\{d^2 \tanh ^{-1}(\phi )^2 \left[b^4 d^2+2 b^2 c_3 \left(d^2-2 c_3\right)\right.\right.\nonumber\\
   &+&d^4 \left(b^2+d^2\right) \tanh ^{-1}(\phi )^6+\tanh ^{-1}(\phi )^2 \left(2 b^4 d^2+3 b^2 \left(d^4-c_3^2\right)+6
   c_3^2 d^2\right)\nonumber\\
   &+&\left.\left.d^2 \tanh ^{-1}(\phi )^4 \left(b^4+4 d^2 \left(b^2+c_3\right)-2 b^2 c_3\right)+4 c_3^3\right]+c_3^2 (c_3-b d) (b d+c_3)\right\}.
\end{eqnarray}
The analytical expressions are quite extensive, but we can analyze the behavior of the functions graphically through Fig. \ref{fig:V-h-phi_model2}. Although the possible range for the scalar field is smaller than in the previous case, now $\phi(r \to \pm \infty) \to \pm 1$, the behavior remains similar to that shown in Fig. \ref{fig:V-h-phi}.

Therefore, we see that even though the chosen method allows us the freedom to select the scalar field, the behavior of the functions for the chosen models is quite similar. Thus, due to the changes introduced by the new parameters, it is now of interest to investigate how these modifications may help mitigate the violation of the energy conditions in this solution.

\section{Energy Conditions}
Wormholes are known to violate energy conditions, such as the NEC. For a wormhole to be stable and allow the passage of matter, it is necessary that exotic matter exists, with properties like negative energy density. Therefore, it is necessary to explore the energy conditions in our wormhole model, investigating whether they are violated.

Besides the NEC mentioned earlier, we can also analyze the strong energy condition (SEC), weak energy condition (WEC), and dominant energy condition (DEC). These conditions are given by the following inequalities \cite{Visser:1995cc}:
\begin{eqnarray}
&&NEC_{1,2}=WEC_{1,2}=SEC_{1,2} 
\Longleftrightarrow \rho+p_{r,t}\geq 0,\label{Econd1} \\
&&SEC_3 \Longleftrightarrow\rho+p_r+2p_t\geq 0,\label{Econd2}\\
&&DEC_{1,2} \Longleftrightarrow \rho-|p_{r,t}|\geq 0 \Longleftrightarrow 
(\rho+p_{r,t}\geq 0) \hbox{ and } (\rho-p_{r,t}\geq 0),\label{Econd3}\\
&&DEC_3=WEC_3 \Longleftrightarrow\rho\geq 0,\label{Econd4}
\end{eqnarray}
where $\rho$, $p_r$, and $p_t$ are the energy density, radial pressure, tangential pressure, respectively. These functions can be identified through the components of the stress-energy tensor as
\begin{equation}
    {T^{\mu}}_{\nu}=\mbox{diag}\left[\rho,\, -p_r,\, -p_t,\, -p_t\right].
\end{equation}
This identification is made through Einstein equations by calculating the Einstein tensor for the line element \eqref{line}.

Calculating the components of the stress-energy tensor, we obtain the following combinations:
\begin{eqnarray}
&&\rho+p_{r}=-\frac{2 b^4 r^2}{\left(r^2+c_3\right)^4}-\frac{2 b^2 \left(d^2 \left(3 r^2-c_3\right)+r^4-3 r^2
   c_3\right)}{\left(d^2+r^2\right) \left(r^2+c_3\right)^3}-\frac{2 d^2}{\left(d^2+r^2\right)^2}, \\
&&\rho+p_{t}=-\frac{2 b^4 r^2}{\left(r^2+c_3\right)^4}+\frac{e^{-\frac{b^2}{r^2+c_3}}-1}{d^2+r^2}+\frac{b^2 \left(d^2
   \left(c_3-3 r^2\right)+r^4+5 r^2 c_3\right)}{\left(d^2+r^2\right) \left(r^2+c_3\right)^3}, \\
&&\rho+p_r+2p_t= 0,\\
&&\rho-p_{r}-\frac{4 b^4 r^2}{\left(r^2+c_3\right)^4}+\frac{2
   \left(e^{-\frac{b^2}{r^2+c_3}}-1\right)}{d^2+r^2}+\frac{2 b^2 \left(d^2 \left(c_3-3 r^2\right)+r^4+5
   r^2 c_3\right)}{\left(d^2+r^2\right) \left(r^2+c_3\right)^3},\\
&&\rho-p_{t}= -\frac{4 b^4 r^2}{\left(r^2+c_3\right)^4}+\frac{e^{-\frac{b^2}{r^2+c_3}}}{d^2+r^2}-\frac{b^2 \left(d^2
   \left(9 r^2-3 c_3\right)+r^4-11 r^2 c_3\right)}{\left(d^2+r^2\right)
   \left(r^2+c_3\right)^3}-\frac{3 d^2+r^2}{\left(d^2+r^2\right)^2}, \\
&&\rho=-\frac{3 b^4 r^2}{\left(r^2+c_3\right)^4}+\frac{2 b^2 \left(d^2 \left(c_3-3 r^2\right)+4 r^2
   c_3\right)}{\left(d^2+r^2\right) \left(r^2+c_3\right)^3}+\frac{\left(d^2+r^2\right)
   e^{-\frac{b^2}{r^2+c_3}}-2 d^2-r^2}{\left(d^2+r^2\right)^2}.
\end{eqnarray}

For consistency, we can check that when $b = 0$ in the combinations above, we have that
\begin{eqnarray}
    \rho + p_r &=& -\frac{2d^2}{(r^2 + d^2)^2} < 0,\,\, \forall\, r\in (-\infty,+\infty),\\
    \rho + p_t &=& 0,\\
    \rho + p_r + 2p_t &=& 0,\\
    \rho - p_r &=& 0,\\
    \rho - p_t &=& -\frac{2d^2}{(r^2 + d^2)^2} < 0,\,\, \forall\, r \in (-\infty,+\infty),
\end{eqnarray}
such that for $b = 0$, the energy conditions are globally violated, which is expected since $b= 0$ recovers the Ellis-Bronnikov case.

\begin{figure}
    \centering
    \includegraphics[width=1\linewidth]{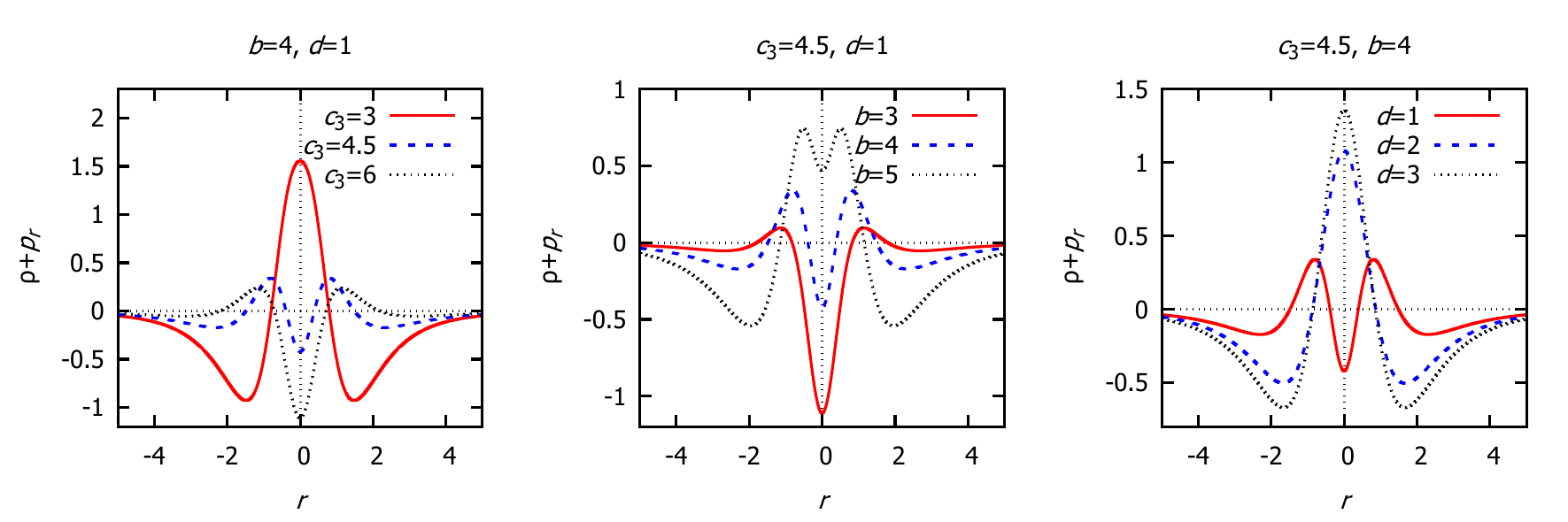}
    \includegraphics[width=1\linewidth]{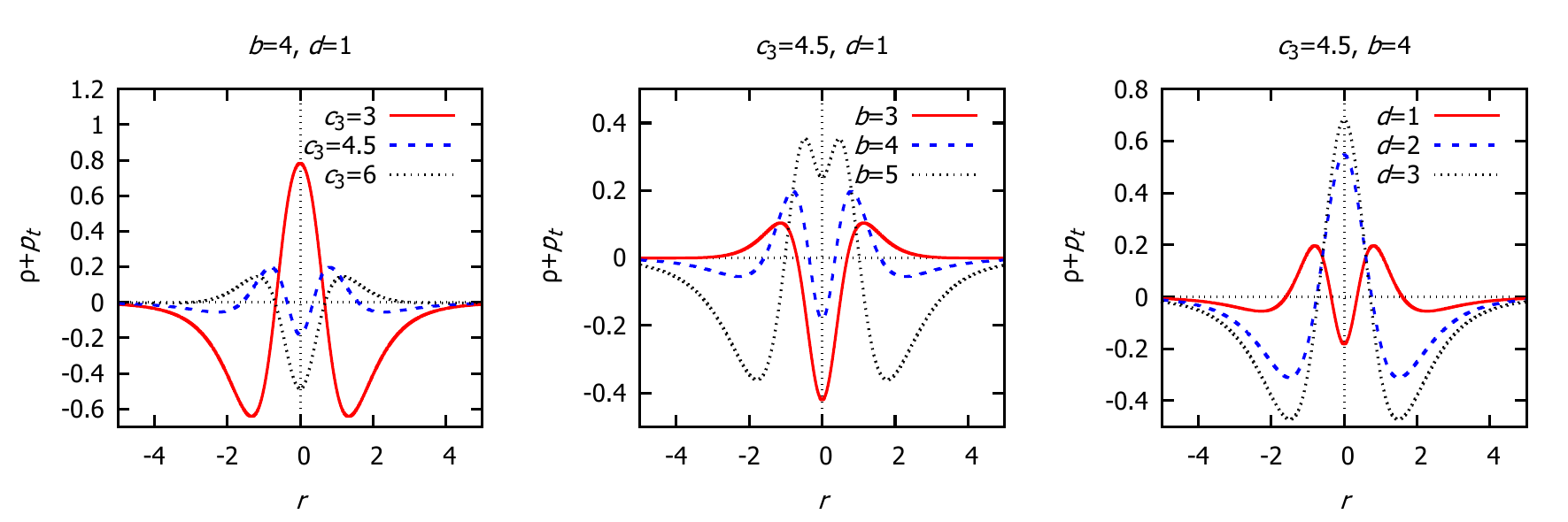}
    \includegraphics[width=1\linewidth]{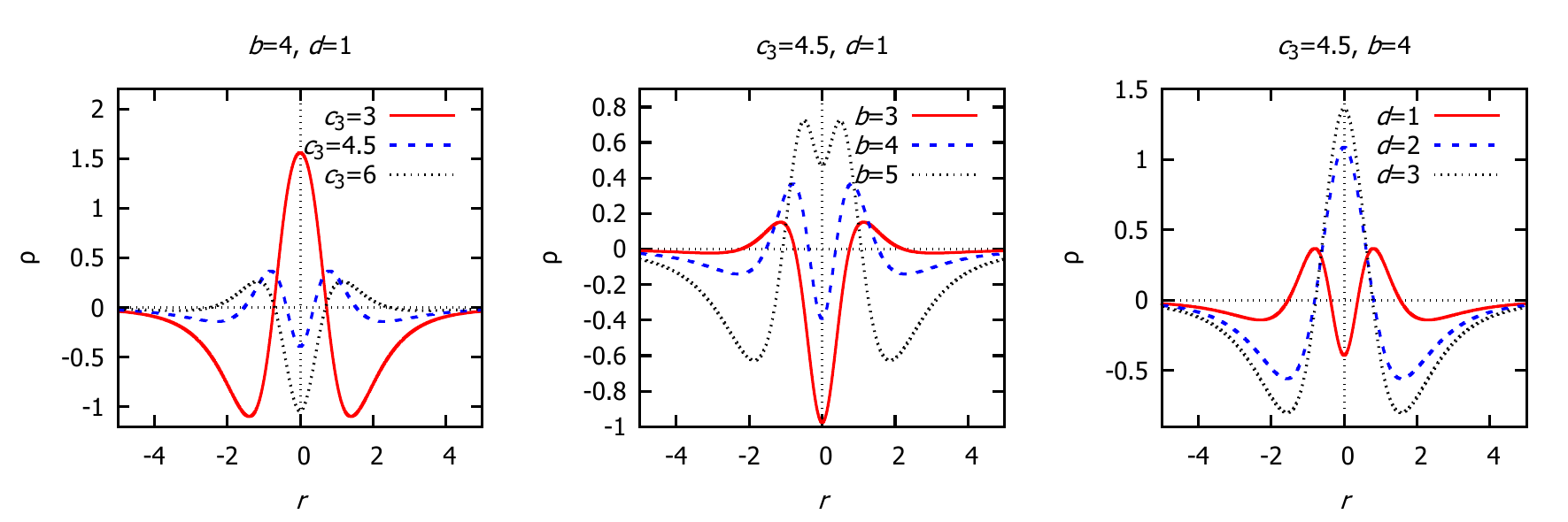}
    \caption{Behavior of the energy density $\rho$ (bottom panel) and the combinations of the pressures $\rho+p_r$ (top panel) and $\rho+p_t$ (middle panel) for different combinations of the wormhole parameters $b, d$ and $c_3$.}
    \label{fig:EC1}
\end{figure}

\begin{figure}
    \centering
    \includegraphics[width=1\linewidth]{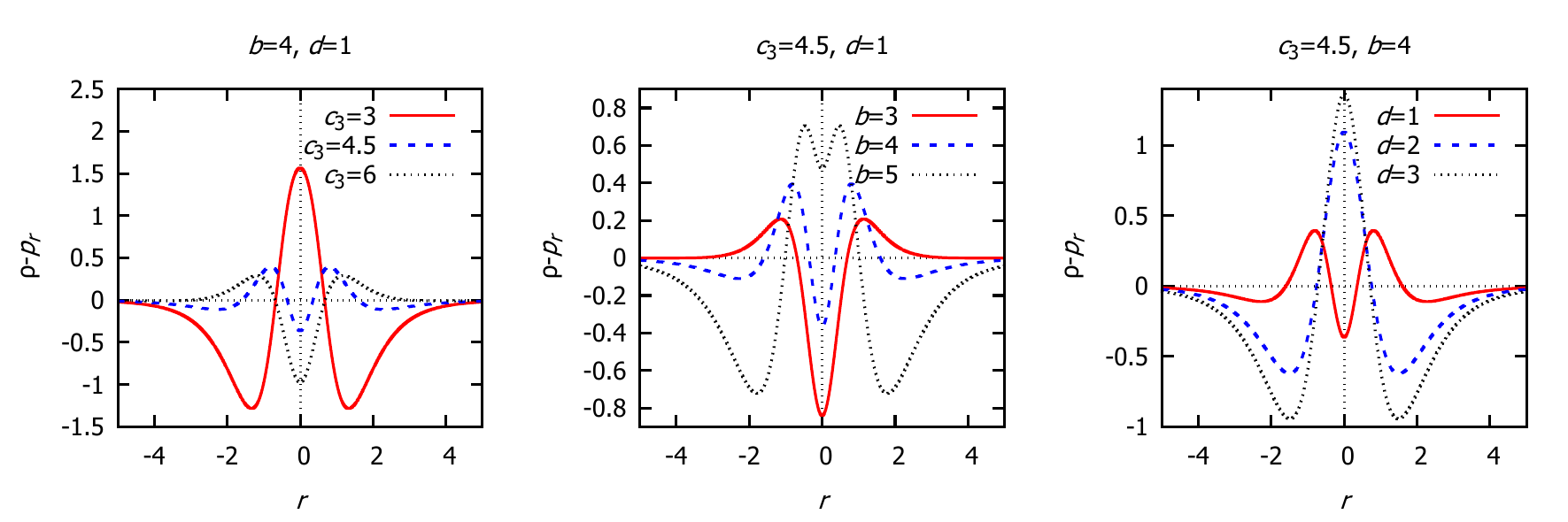}
    \includegraphics[width=1\linewidth]{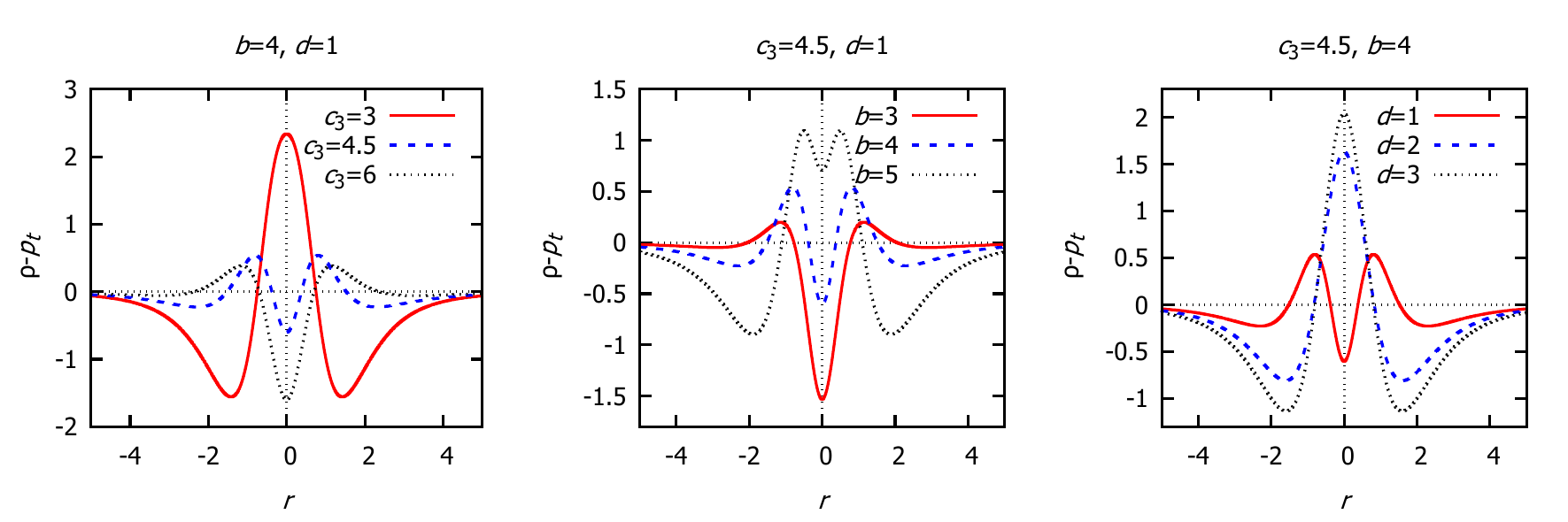}
    \caption{Behavior of the combinations of the pressures $\rho-p_r$ (top panel) and $\rho-p_t$ (bottom panel) for different combinations of the wormhole parameters $b, d$ and $c_3$.}
    \label{fig:EC2}
\end{figure}

For $b \neq 0$, the only condition that is identically satisfied is $SEC_3$. The other conditions need to be further analyzed by checking whether the combinations above are positive or not. In Figs. \ref{fig:EC1} and \ref{fig:EC2}, we analyze the behavior of the energy density and its combinations with the pressures. We verify that, regardless of the chosen parameters, there are always regions where the energy conditions are violated. Depending on the choice of parameters, it is possible to satisfy all energy conditions in the central region, $r\rightarrow 0$. For regions farther from the wormhole, the energy conditions are clearly violated. This is consistent with the results we obtained for the scalar field, which, in more distant regions, would necessarily be phantom.

Thus, we observe that our model exhibits both drawbacks and advantages compared to other wormhole models, such as the Ellis-Bronnikov model. Our model no longer guarantees that certain inequalities, which were previously always satisfied, remain valid throughout all regions of spacetime. However, we can ensure that, for certain parameter choices, all energy conditions are satisfied in at least some regions of spacetime. The relaxation in the violation of these energy conditions may imply fewer instabilities, repulsive gravity, or issues with non-causality related to the solution.

\section{Conclusions}

In this work, we study the general characteristics of wormholes geometries with multiple throats/anti-throats, focus on the fields that can serve as sources for this geometry in the context of general relativity. We consider a wormhole model with a trivial redshift function modifying the black bounce metric function given in reference \cite{Rodrigues:2022mdm}. The presence of multiple throats/anti-throats can be observed analyzing the minima and maxima of the area of the sphere in this coordinate system, or, from a geometric point of view, analyzing the embedded diagrams presented in Fig. \ref{diagram}. Depending on the choice of the solution parameters, it is possible to observe the presence of a single throat or even multiple throats/antithroats. Furthermore, considering geodesic trajectories in this spacetime, we find that massive particles follow free paths radially, unlike in the Schwarzschild case. Considering non-radial trajectories, we derived the equation \eqref{angleradioeq}, that describes such paths. This equation can be used to determine the appearance of this object to distant observers, as well as the shadows of this wormhole solution. However, due to the extremely complicated form of the $\Sigma(r)$ function, this becomes highly non-trivial and might be addressed in a future work.

To determine the field sources of the geometry, we started with an action of gravity, in this case described by general relativity, minimally coupled to a scalar field with a non-trivial potential and a NED Lagrangian. We derived the equations of motion and expressed the electromagnetic Lagrangian $L(F(r))$, its derivative $L_F(F(r))$ the scalar potential $V(\phi(r))$, and the coupling function $h(\phi(r))$ in terms of the wormhole metric function $\Sigma(r)$ and the scalar field $\phi(r)$. By manipulating the equations, we demonstrated that the functions $L$, $L_F$, and $V$, expressed as functions of the radial coordinate $r$, are independent of the specific form of the scalar field. In other words, different choices for $\phi$ will yield the same $L(r)$, $L_F(r)$, and $V(r)$. 

We tested two widely studied models for $\phi$ in the literature, where the scalar field takes the functional form of an arctangent, $\phi(r) = \arctan(r/d)$, and a hyperbolic tangent, $\phi(r) = \tanh(r/d)$. With this, we explicitly determined the functional forms of $V(\phi)$ and $h(\phi)$ for both cases. We observed that $h(\phi)$ transits between positive and negative values, indicating regions where the scalar field alternates between standard and phantom behaviors. The scalar potential 
$V(\phi)$ decreases to zero asymptotically, while the coupling $h(\phi)$ approaches a constant, ensuring a smooth behavior at large radial distances. We can further state that the general behavior of these functions does not significantly change for the two chosen scalar field models. We also determined the function $L(F)$ analytically as functions of $F$. In terms of $r$, we have the following asymptotic behaviors, both as $r \to \infty$, and as $r \to 0$, respectively, 
\begin{eqnarray}
    L(r) &\to& \frac{-b^4 + 2 b^2 c_3 - 2 b^2 d^2}{r^6} + O\left(\frac{1}{r^7}\right),\\
    L(r) &\to& constant.
\end{eqnarray}

The analysis of the energy conditions revealed that, although the wormhole geometry generally violates all energy conditions, there are parameter choices where these conditions can be satisfied near the wormhole's throat. However, such violations are inevitable at larger radial distances, consistent with the exotic matter requirements of traversable wormholes in general relativity.

In conclusion, the results demonstrate that the combination of a scalar field and nonlinear electrodynamics is capable of generating physically consistent wormhole solutions with multiple throats/anti-throats with well-defined fields and potentials. The framework allows flexibility in adjusting parameters to minimize energy condition violations in specific regions, making it a valuable approach for exploring realistic wormhole models.

In addition to the characteristics studied in this article, several physical properties can still be explored in future works. Typically, wormholes can exhibit echoes \cite{Magalhaes:2023xya,Cardoso:2016rao,Churilova:2021tgn}; thus, we can analyze the echoes produced by our wormhole model and compare these results with those of other known models. Furthermore, our solution can be compared with other models in terms of its optical appearance or through the deflection of light caused by these wormholes \cite{Shaikh:2019jfr,Jusufi:2017vta,Bronnikov:2018nub}.

\section*{Acknowledgments}
\hspace{0.5cm} The authors would like to thank Conselho Nacional de Desenvolvimento Cient\'{i}fico e Tecnol\'ogico (CNPq) and Funda\c c\~ao Cearense de Apoio ao Desenvolvimento Cient\'ifico e Tecnol\'ogico (FUNCAP) for partial financial support. CRM and GA thanks the Conselho Nacional de Desenvolvimento Cient\'{i}fico e Tecnol\'{o}gico (CNPq), Grants no. 308268/2021-6 and 315568/2021-6. This work is also supported by the Spanish grant Ref.~PID2020-117301GA-I00 (DS-CG) funded by MCIN/AEI/10.13039/501100011033 (``ERDF A way of making Europe" and ``PGC Generaci\'on de Conocimiento"), and by the Department of Education, Junta de Castilla y Le\'on (Spain) and FEDER Funds (Ref. CLU-2023-1-05).


\bibliography{Ref.bib}
\end{document}